\documentclass[twocolumn]{aastex63}
\pdfoutput=1 
\usepackage[T1]{fontenc}
\usepackage{CJK}
\usepackage{amsmath,amstext}
\usepackage{apjfonts} 
\usepackage[figure,figure*]{hypcap}
\usepackage{microtype}
\usepackage{tablefootnote}
\usepackage{booktabs}  
\usepackage{longtable}
\usepackage{hyperref}

\begin{document}
\begin{CJK*}{UTF8}{gbsn}
\shortauthors{Eftekhari et al.}

\shorttitle{An X-ray Census of Fast Radio Burst Host Galaxies}

\uppercase{\title{\normalfont An X-ray Census of Fast Radio Burst Host Galaxies: Constraints on AGN and X-ray Counterparts}}
\newcommand{\NU}{\affiliation{Center for Interdisciplinary Exploration and Research in Astrophysics (CIERA) and Department of Physics and Astronomy, Northwestern University, Evanston, IL 60208, USA}}

\newcommand{\Swinburne}{\affiliation{Centre for Astrophysics and Supercomputing, Swinburne University of Technology, John St, Hawthorn, VIC 3122, Australia}}

\newcommand{\ASTRON}{\affiliation{ASTRON, Netherlands Institute for Radio Astronomy, Oude Hoogeveensedijk 4, 7991 PD Dwingeloo, The Netherlands}}

\newcommand{\VLBI}{\affiliation{Joint institute for VLBI ERIC, Oude Hoogeveensedijk 4, 7991 PD Dwingeloo, The Netherlands
}}

\newcommand{\Anton}{\affiliation{Anton Pannekoek Institute for Astronomy, University of Amsterdam, Science Park 904, 1098 XH, Amsterdam, The Netherlands}}

\newcommand{\UCSC}{\affiliation{Department of Astronomy and Astrophysics, University of California, Santa Cruz, CA 95064, USA}}

\newcommand{\STScI}{\affiliation{Space Telescope Science Institute, Baltimore, MD 21218, USA}}

\newcommand{\JHU}{\affiliation{Department of Physics and Astronomy, Johns Hopkins University, Baltimore, MD 21218, USA}}

\newcommand{\DAWN}{\affiliation{Cosmic Dawn Center (DAWN), Denmark}}

\newcommand{\Bohr}{\affiliation{Niels Bohr Institute, University of Copenhagen, Jagtvej 128, DK-2200 Copenhagen N, Denmark}}

\newcommand{\PSU}{\affiliation{Department of Astronomy \& Astrophysics, The Pennsylvania State University, University Park, PA 16802, USA}}

\newcommand{\DS}{\affiliation{Institute for Computational \& Data Sciences, The Pennsylvania State University, University Park, PA 16802, USA}}

\newcommand{\GC}{\affiliation{Institute for Gravitation and the Cosmos, The Pennsylvania State University, University Park, PA 16802, USA}}

\newcommand{\CSIRO}{\affiliation{CSIRO, Space and Astronomy, PO Box 76, Epping, NSW 1710, Australia}}

\newcommand{\Fisica}{\affiliation{Instituto de F\'isica, Pontificia Universidad Cat\'olica de Valpara\'iso, Casilla 4059, Valpara\'iso, Chile}}

\newcommand{\Macquarie}{\affiliation{Department of Physics \& Astronomy, Macquarie University, Sydney, NSW 2109, Australia}}

\newcommand{\CfAMacquarie}{\affiliation{Macquarie University, Research Centre for Astronomy, Astrophysics \& Astrophotonics, Sydney, NSW 2109, Australia}}

\newcommand{\CCA}{\affil{Center for Computational Astrophysics, Flatiron Institute, 162 5th Ave, New York, NY 10010, USA} }

\newcommand{\THEA}{\affiliation{Theoretical High Energy Astrophysics (THEA) Group, Columbia University, New York, New York 10027, USA}}

\newcommand{\Berkeley}
{\affiliation{Astronomy Department and Theoretical Astrophysics Center, University of California, Berkeley, Berkeley, CA 94720, USA}}

\newcommand{\Columbia}{\affiliation{Department of Astronomy, Columbia University, New York, NY 10027, USA}}

\newcommand{\IPMU}{\affiliation{Kavli Institute for the Physics and Mathematics of the Universe (Kavli IPMU), 5-1-5 Kashiwanoha, Kashiwa, 277-8583, Japan}}

\newcommand{\NAOJ}{\affiliation{Division of Science, National Astronomical Observatory of Japan, 2-21-1 Osawa, Mitaka, Tokyo 181-8588, Japan}}

\newcommand{\MQ}{\affiliation{School of Mathematical and Physical Sciences, Macquarie University, NSW 2109, Australia}}

\newcommand{\ASTRC}{\affiliation{Astrophysics and Space Technologies Research Centre, Macquarie University, Sydney, NSW 2109, Australia}}

\newcommand{\ESA}{\affiliation{European Space Agency (ESA), European Space Astronomy Centre (ESAC), Camino Bajo del Castillo s/n, 28692 Villanueva de la Cañada, Madrid, Spain}}

\newcommand{\McGill}{\affiliation{Department of Physics, McGill University, 3600 rue University, Montr\'eal, QC H3A 2T8, Canada}}

\newcommand{\TSI}{\affiliation{Trottier Space Institute, McGill University, 3550 rue University, Montr\'eal, QC H3A 2A7, Canada}}

\newcommand{\DUN}{\affiliation{Dunlap Institute for Astronomy \& Astrophysics, University of Toronto, 50 St. George Street, Toronto,
ON M5S 3H4, Canada}}
\author[0000-0003-0307-9984]{T. Eftekhari}
\altaffiliation{NASA Einstein Fellow}
\NU

\author[0000-0002-7374-935X]{W. Fong}
\NU

\author[0000-0002-5025-4645]{A. C. Gordon}
\NU

\author[0000-0002-5519-9550]{N. Sridhar}
\Columbia
\THEA

\author[0000-0002-5740-7747]{C. D. Kilpatrick}
\NU

\author[0000-0003-3460-506X]{S. Bhandari}
\altaffiliation{Veni Fellow}
\ASTRON
\VLBI
\Anton
\CSIRO

\author[0000-0001-9434-3837]{A. T. Deller}
\Swinburne

\author[0000-0002-9363-8606]{Y. Dong (董雨欣)}
\NU

\author[0000-0003-3937-0618]{A. Rouco Escorial}
\altaffiliation{ESA Reseach Fellow}
\ESA

\author[0000-0002-9389-7413]{K. E. Heintz}
\DAWN
\Bohr

\author[0000-0001-6755-1315]{J. Leja}
\PSU
\DS
\GC

\author[0000-0001-8405-2649]{B. Margalit}
\Berkeley

\author[0000-0002-4670-7509]{B. D. Metzger}
\Columbia
\THEA
\CCA

\author[0000-0002-8912-0732]{A. B. Pearlman}
\McGill
\TSI

\author[0000-0002-7738-6875]{J. X. Prochaska}
\UCSC
\IPMU
\NAOJ

\author[0000-0003-4501-8100]{S. D. Ryder}
\MQ
\ASTRC

\author[0000-0002-7374-7119]{P. Scholz}
\DUN

\author[0000-0002-7285-6348]{R. M. Shannon}
\Swinburne

\author[0000-0002-1883-4252]{N. Tejos}
\Fisica

\begin{abstract}
We present the first X-ray census of fast radio burst (FRB) host galaxies to conduct the deepest search for AGN and X-ray counterparts to date. Our sample includes seven well-localized FRBs with unambiguous host associations and existing deep \textit{Chandra} observations, including two events for which we present new observations. We find evidence for AGN in two FRB host galaxies based on the presence of X-ray emission coincident with their centers, including the detection of a luminous ($L_X\approx\,5\times\,10^{42}\,\rm\,erg\,s^{-1}$) X-ray source at the nucleus of FRB\,20190608B's host, for which we infer an SMBH mass of $\rm{M_{BH}\sim\,10^{8}\,M_{\odot}}$ and an Eddington ratio $\rm{L_{bol}/L_{Edd}\approx\,0.02}$, characteristic of geometrically thin disks in Seyfert galaxies. We also report nebular emission line fluxes for 24 highly secure FRB hosts (including 10 hosts for the first time), and assess their placement on a BPT diagram, finding that FRB hosts trace the underlying galaxy population. We further find that the hosts of repeating FRBs are not confined to the star-forming locus, contrary to previous findings. 
Finally, we place constraints on associated X-ray counterparts to FRBs in the context of ultraluminous X-ray sources (ULXs), and find that existing X-ray limits for FRBs rule out ULXs brighter than $L_X\gtrsim\,10^{40}\,\rm\,erg\,s^{-1}$. Leveraging the CHIME/FRB catalog and existing ULX catalogs, we search for spatially coincident ULX-FRB pairs. We identify a total of 28 ULXs spatially coincident with the localization regions for 17 FRBs, but find that the DM-inferred redshifts for the FRBs are inconsistent with the ULX redshifts, disfavoring an association between these specific ULX-FRB pairs.
\end{abstract}


\section{Introduction}
\label{sec:intro}

Fast radio bursts (FRBs) are bright, millisecond duration flares of coherent radio emission \citep{Lorimer2007,Thornton2013} whose large dispersion measures (DMs) point to an extragalactic origin. Despite over 500 FRBs published in the literature to date (e.g., \citealt{Macquart2020,CHIME2020,CHIME_repeaters2023,Law2023}), the sources responsible for producing FRBs are highly debated, in large part due to the dearth of well-localized events \citep{Eftekhari2017} as well as the fact that some FRBs have been known to repeat (so-called ``repeaters''; \citealt{Spitler2016}) while others appear as one-off events \citep{Shannon2018,CHIME2021}. Although repeating and apparently non-repeating FRBs exhibit a number of distinctions in their spectro-temporal properties \citep{Pleunis2021} --- and potentially their host galaxy properties \citep{Gordon2023} --- analyses of the FRB population cannot exclude the possibility that all FRBs intrinsically repeat \citep{James2023}.

To reconcile this apparent diversity in burst properties and environments, a wide range of theoretical models have been put forth for FRB progenitors, ranging from compact object mergers \citep{Margalit2019, Sridhar+21} and flares from highly magnetized stars \citep{Margalit2018} to the super-Eddington accretion onto a neutron star or black hole \citep{Sridhar2021, Sridhar+23b}, or even more exotic phenomena invoking cosmic strings \citep{Costa2018} and primordial black holes \citep{Abramowicz2018}. At present, prevailing theoretical models invoke magnetars, whose extremely large magnetic fields are capable of storing enormous magnetospheric energies released in FRB flashes. Indeed, the discovery of a luminous, millisecond-duration burst from the known Galactic magnetar SGR 1935+2154 \citep{CHIME2020,Bochenek2020} supports a connection between extragalactic magnetars and FRBs \citep{Margalit2020}. Recent studies suggest that magnetars formed via multiple formation channels --- and not strictly through core-collapse supernovae -- likely contribute to the known FRB population (e.g., \citealt{Margalit2019,Kirsten2022,Pelliciari2023}).

Technical upgrades to a number of FRB experiments in recent years have facilitated a small, but growing number of (sub-)arcsecond localizations and robust host galaxy identifications for approximately three dozen events (e.g., \citealt{Bhardwaj2021,Kirsten2022,Ryder2022,Gordon2023,Ravi2023,Law2023,Sharma2023}). The first population studies demonstrate that FRB hosts span a wide range of stellar masses and star formation rates and trace the underlying population of field galaxies \citep{Heintz2020,Bhandari2022,Gordon2023}. 

These early host galaxy compilations have also revealed that a non-negligible fraction of FRB hosts occupy a distinct region of the Baldwin-Phillips-Terlevich (BPT; \citealt{Phillips1981}) diagram, pointing to emission line ratios in excess of typical star forming galaxies (\citealt{Heintz2020,Bhandari2022,Ibik2023}). In particular, non-repeating FRBs seem to indicate a preference for a sub-population of low ionization nuclear emission line region (LINER) galaxies, while repeating FRBs lie exclusively along the star-forming branch. This distinction suggests that non-repeating FRBs may be preferentially located in environments with enhanced levels of photoionization. 

The nature of LINER galaxies, however, and the source of their ionizing photons have been under debate since their discovery \citep{Heckman1980}. Existing theories associate LINERs with low luminosity active galactic nuclei (AGN) or stellar ionization from evolved post-asymptotic giant branch (post-AGB) stars \citep{Singh2013}. For FRB hosts, the latter scenario would implicate older stellar populations and hence FRB progenitors formed through channels with large average delay times between star formation and FRB source formation (e.g., short-duration gamma-ray bursts). Conversely, if the hard radiation fields observed in some FRB hosts are driven by central AGN, pointing to an elevated fraction of AGN in FRB hosts, this would lend support for theories that presuppose an enhanced rate of compact objects formed in AGN accretion disks \citep{Cantiello2021,Perna2021a,Perna2021b,Jermyn2021} or FRB progenitor models that invoke AGN \citep{Bing2017}. Indeed, several of these theories have been put forth for relativistic transients, including both long- and short-duration gamma-ray bursts, but could be applied here given the higher stellar densities in AGN disks in general. Moreover, with only a small sample of FRB hosts that have been analyzed in this context \citep{Heintz2020,Bhandari2022}, a larger study is warranted to quantify the true occurrence rate of AGN (and LINER galaxies) in FRB hosts.

Concurrent to discerning the origin of the hard radiation fields prevalent in FRB hosts, X-ray observations can uniquely probe high-energy counterparts to FRBs. Constraints on periodicity in the burst rate from two repeating FRBs \citep{Rajwade2020,CHIME2020_periodic} have prompted theories that they are powered by relativistic flares along precessing jets associated with accretion onto compact objects \citep{Sridhar2021}. The observed (ms) durations and the extreme luminosities of typical FRBs, as well as the properties of their host galaxies, suggest that FRBs may be produced by stellar-mass ultraluminous X-ray sources (ULXs) powered by the super-Eddington accretion onto a compact object \citep{Sridhar2021}. Within this framework, bright ULX-like X-ray emission should be associated with at least some FRBs. While deep limits on X-ray counterparts exist for a small number of well-localized FRB sources (e.g., \citealt{Scholz2020,Kirsten2022,Pearlman2023}) and targeted searches for FRBs from nearby galaxies hosting ULXs have been conducted \citep{Pelliciari2023}, there have been no concerted efforts thus far to search for ULXs coincident with the localization regions of the hundreds of FRBs detected across the sky.

Here we report the X-ray properties for all well-localized FRBs with existing \textit{Chandra} X-ray observations (including newly obtained data for two events) to probe the presence of both AGN in FRB hosts and X-ray counterparts at the FRB locations. Our sample consists of seven FRBs, including five repeating sources. In Section~\ref{sec:obs}, we present our observations and data reduction, which includes new Karl G. Jansky Very Large Array (VLA) observations for two FRBs. In Section~\ref{sec:AGN}, we combine radio and X-ray observations of our sample of hosts to constrain the multiwavelength properties of AGN in FRB host galaxies. We additionally present an updated BPT analysis for a uniformly-modeled sample of 22 FRB host galaxies (in addition to two hosts from the literature). We leverage the X-ray observations to place limits on associated X-ray counterparts to FRBs in Section~\ref{sec:ulx}, and we further explore a possible connection between FRBs and ULXs by searching for FRBs beyond the original sample that are spatially coincident with catalogued ULXs. We summarize our results in Section~\ref{ref:conclusions}. Throughout the paper, we use the latest Planck cosmological parameters for a flat $\Lambda$CDM universe, with $H_{0}$ = 67.66 km s$^{-1}$ Mpc$^{-1}$, $\Omega_m = 0.310$, and $\Omega_{\lambda} = 0.690$ \citep{Planck2020}.

\section{Observations}\label{sec:obs}

\subsection{Sample Selection}

Our sample of FRB hosts includes all well-localized ($\sim$ a few to hundreds of milli-arcseconds) events with unambiguous host associations (a probabilistic association of transients to their hosts (PATH; \citealt{Aggarwal2021}) posterior probability of $\gtrsim 90\%$) and existing \textit{Chandra} data. The resulting sample spans a redshift range of $z \sim 0.0 - 0.2418$\footnote{The $z=0$ source corresponds to FRB\,20200120E located in the nearby M81 galaxy at $d=3.6$ Mpc \citep{Kirsten2022}.} and includes five FRBs with previously published X-ray data which we re-analyze here and two for which we present new \textit{Chandra} observations (see Table~\ref{tab:xrayfrbs}). The FRB hosts in our sample span a wide range in both stellar masses ($\sim \rm log(M_{*}/M_{\odot}) = 8.14 - 10.84$) and star formation rates ($\sim 0.04 - 7.03 \ \rm M_{\odot} \ yr^{-1}$), similar to the general population of FRB hosts \citep{Gordon2023}, although are biased towards repeating events (five out of seven FRBs).

\subsection{X-ray Detection of the AGN in the Host Galaxy of FRB\,20190608B}

We obtained X-ray observations with the {\it Chandra Advanced CCD Imaging Spectrometer} (ACIS-S; Obs ID: 25251; PI: Eftekhari) of the field of FRB\,20190608B (a non-repeating FRB discovered and localized by the Australian Square Kilometre Array Pathfinder (ASKAP) \citep{Macquart2020}) on UT 2021 December 6 with an exposure time of 19.8 ks. We analyzed the data using the \texttt{CIAO} software package (v4.13) and followed standard ACIS data filtering. 

\begin{figure*}
\includegraphics[width=0.94\columnwidth]
{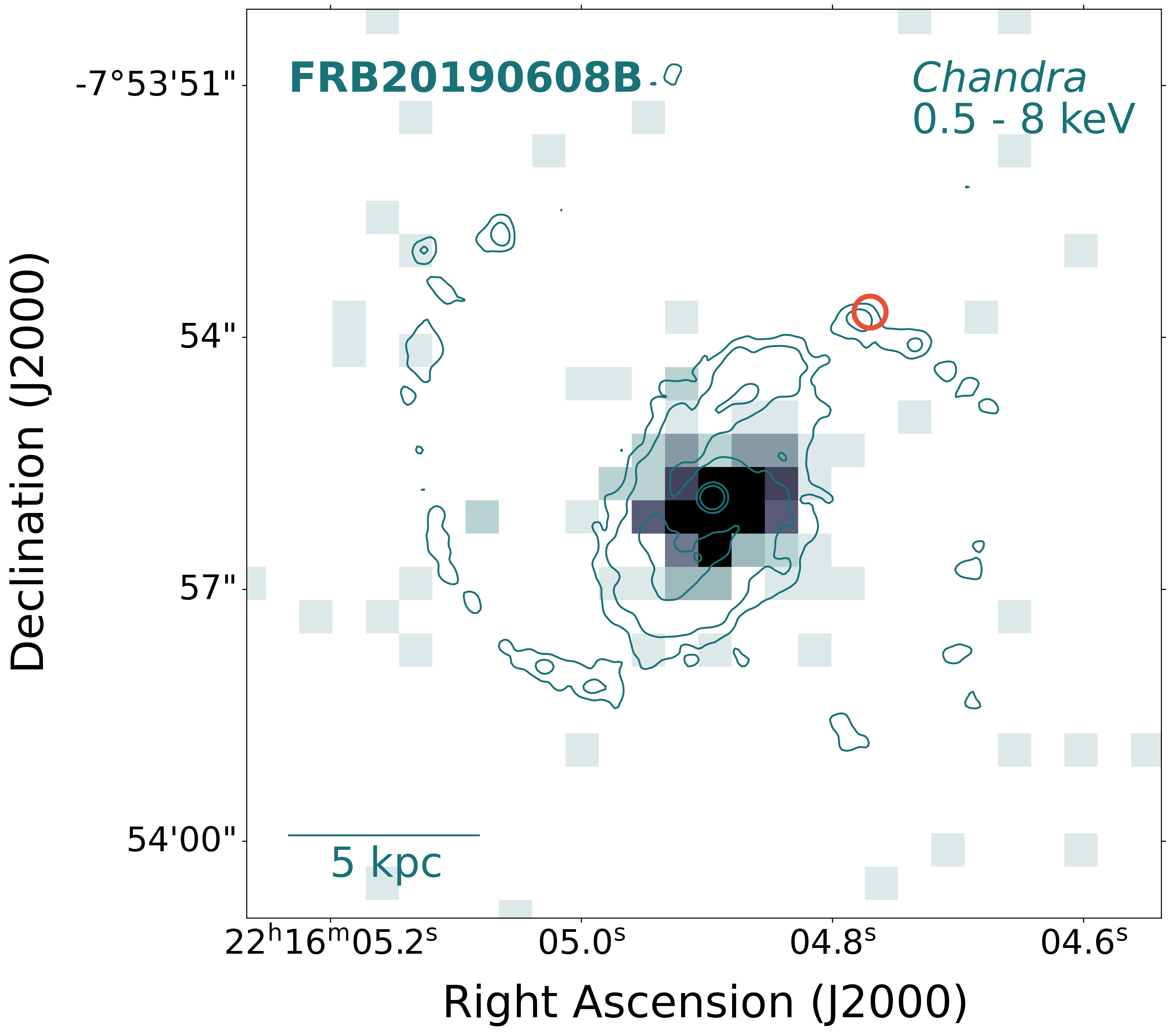}
\hspace{0.2cm}
\includegraphics[width=1.01\columnwidth]{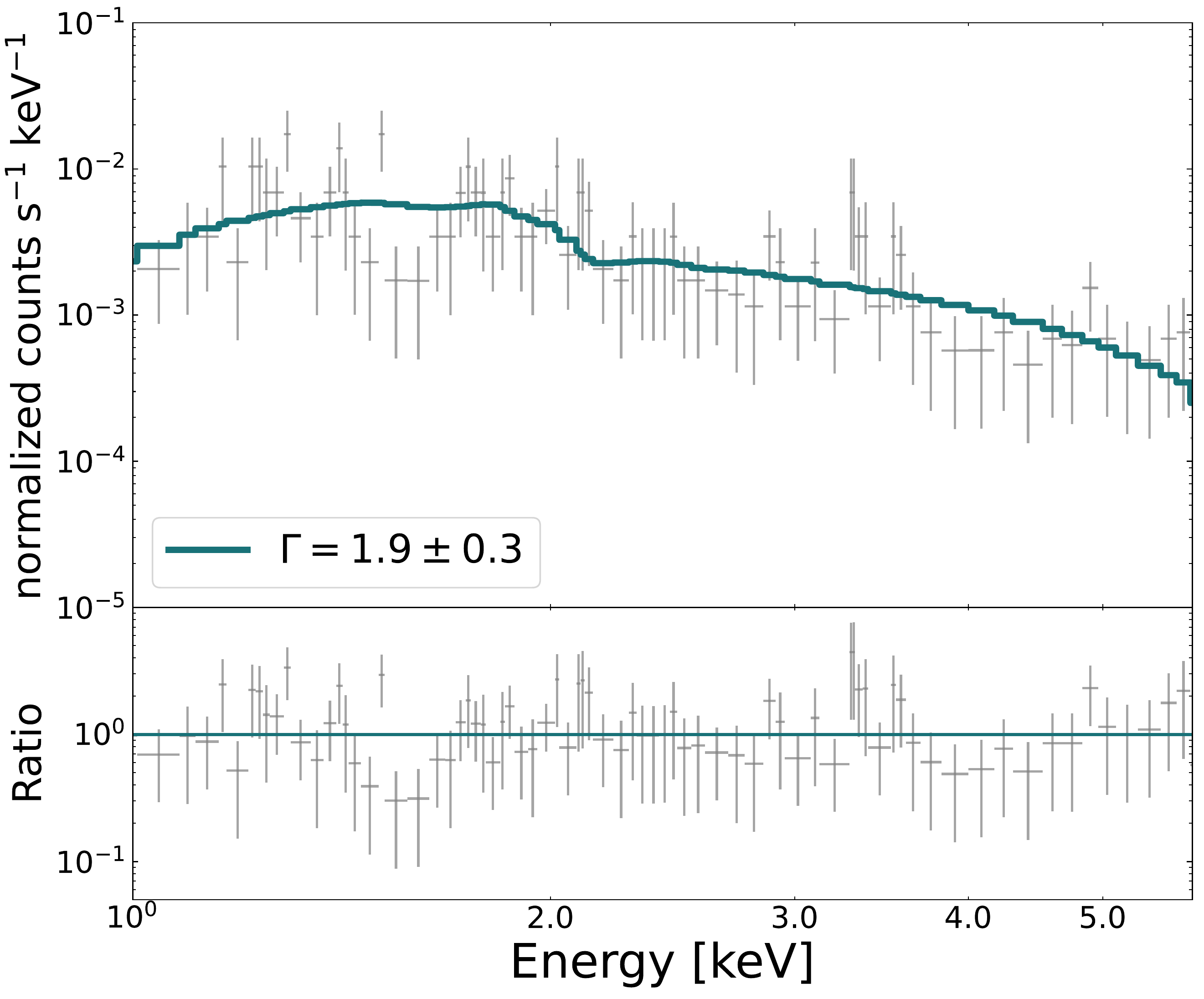}
\caption{\textbf{Left:} \textit{Chandra} X-ray detection (0.5 - 8 keV) of the host galaxy of FRB\,20190608B. Contours from the Gaussian-smoothed \textit{HST} (F300X) data \citep{Chittidi2021} are overlaid in green. The red circle corresponds to the $1\sigma$ positional uncertainty on the FRB position \citep{Day2021}. \textbf{Right:} X-ray spectrum (grey points) of the AGN in the host galaxy of FRB\,20190608B, modeled with an absorbed power law (green curve) with a best fit power law index $\Gamma = 1.9\pm 0.3$.}
\label{fig:frb190608_image}
\end{figure*}

We detect a bright X-ray source offset from the FRB position by $\approx 1.6\arcsec$ with high significance ($71\sigma$) using a blind search with \texttt{wavdetect} (Figure~\ref{fig:frb190608_image}). The source is detected in an elliptical region of size $2.8\arcsec\times2.5\arcsec$ (PA $= 32^\circ$) centered at RA = 22h16m04.858s, Decl = -7d53m56.297s. To estimate the background, we create a $35\arcsec$ radius region near the center of the chip and free from any obvious sources. We find a net source count rate of $(108.3 \pm 7.5)\times 10^{-4} \ \rm cts \ s^{-1}$ (0.5 - 8 keV). Given the spatial coincidence of the source with the host galaxy center and the non-extended nature of the X-ray emission, we consider the source a putative AGN.

We re-bin the spectrum across the $2.8\arcsec\times2.5\arcsec$ source region using 2 counts per bin and fit the data using an absorbed power-law model (\texttt{tbabs$\ast$ztbabs$\ast$pow} in \texttt{Xspec}; \citealt{Arnaud1996}) with a fixed Galactic absorption column density $N_{\rm H,MW} = 5.9 \times 10^{20} \ \rm cm^{-2}$ \citep{Willingale2013}, solar abundances from \citet{Asplund2009}, and use the maximum likelihood C-statistic for Poisson statistics (\texttt{c-stat}; \citealt{Cash1979}). We find a photon index of $\Gamma = 1.9 \pm 0.3$ (68\% confidence) and an intrinsic column density $N_{\rm H} = (4.7\pm 0.4) \times 10^{21} \ \rm cm^{-2}$ for the AGN in the host galaxy. The model fit is shown in the right panel of Figure~\ref{fig:frb190608_image}. To measure the quality of the fit, we use the Cramer-von Mises (CvM) test statistic (\texttt{statistic test cvm}; \citealt{Cramer1928, Mises1928}) and \texttt{XSPEC}'s ``goodness'' simulations to simulate realizations of the power-law and determine what percentage of simulations have test statistics less than that of the data. A value of $\sim 50\%$ indicates that the observed spectrum was produced by the model. We find that $43\%$ of the realizations have a lower CvM statistic than the data, indicating the model is a good fit. Applying these spectral parameters to the net count rate and employing \texttt{cflux}, we find an unabsorbed X-ray flux $F_X = (1.4\pm 0.2) \times 10^{-13} \ \rm erg \ cm^{-2} \ s^{-1}$ ($2-10$ keV), corresponding to an X-ray luminosity $L_{\rm X} = (5.4\pm 0.8) \times 10^{42} \ \rm erg \ s^{-1}$ at the redshift of the FRB host, $z=0.1178$ (see Table~\ref{tab:xrayfrbs}).

To determine the location of the X-ray source relative to the host galaxy center, any substructure within the host galaxy, and the position of FRB\,20190608B, we use the Hubble Space Telescope (\textit{HST}) F300X image of the host galaxy \citep{Chittidi2021} which enables a precise host centroid. Since there are not enough sources in common between the \textit{HST} and \textit{Chandra} images, we use a Dark Energy Camera Legacy Survey image (DECaLS, DR10; \citealt{Dey2019}) to obtain a common absolute astrometric frame. First, we perform relative astrometry between the \textit{HST} image using \texttt{Source Extractor} \citep{Bertin1996} and point sources in common. We then register the \textit{Chandra} field of view to the DECaLS astrometric system using sources in common. With both the \textit{HST} and \textit{Chandra} images tied to the same astrometric frame, we find that the X-ray source is coincident with the host galaxy center (Figure~\ref{fig:frb190608_image}).

The observed properties of the X-ray source, namely its spatial coincidence with the host galaxy center, luminous X-ray emission at the level of $L_{\rm X} = 5.4 \times 10^{42} \ \rm erg \ s^{-1}$, and an X-ray spectrum that is well-characterized by an absorbed power-law, indicate that the X-ray source is powered by an AGN. We note that we do not detect any counts at the FRB position, corresponding to a $3\sigma$ upper limit on the absorbed X-ray flux of $F_X < 3 \times 10^{-15} \ \rm erg \ cm^{-2} \ s^{-1}$ ($2-10$ keV).

\begin{deluxetable*}{lccccccc}
\tablecolumns{5}
\caption{X-ray and Radio Observations of the Host Galaxies of Localized FRBs}
\tablehead{
\colhead{FRB} &
\colhead{$z$} &
\colhead{Average Observation Epoch} &
\colhead{Exposure Time} &
\colhead{log($L_{\rm 2 - 10 \ keV}$)} &
\colhead{log($L_{\rm 5 \ GHz}$)} &
\colhead{$N_H$} & 
\colhead{References}\\
\colhead{} & 
\colhead{} & 
\colhead{(MJD)} &
\colhead{(ks)} &
\colhead{($\rm erg \ s^{-1}$)} &
\colhead{($\rm erg \ s^{-1}$)} &
\colhead{($\rm 10^{22} \ cm^{-2}$)} &
\colhead{}
}  
\startdata
FRB\,20121102A & 0.1927  & 2016 Nov 30 & 99.9$^{a}$ & $<41.2$ & 39.1 & 0.63 & 1, 2, 3, 4 \\ 
FRB\,20180916B & 0.0337 & 2019 Dec 10 & 32.7$^{b}$ & $<40.0$  & $<36.1$ & 0.61 & 5, 6\\ 
FRB\,20190520B & 0.2418 & 2020 Sep 14 & 14.9 & $<41.9$ &  39.1 & 0.26 & 7,8 \\ 
FRB\,20190608B &  0.1178 & 2021 Dec 06& 19.8 & 42.7$^{+0.06}_{-0.06}$ & $38.1$ & 0.06 & This work; 9\\ 
FRB\,20200120E & 0.0008 & 2010 Jan 24 & $1205.3^{c}$ & 40.4$^{+0.003}_{-0.003}$ & 37.0 & 0.04 & 10, 11, 12 \\
FRB\,20200430A & 0.1608 &  2022 Apr 24 & 34.8$^{d}$ & $<41.4$ & $<37.7$ & 0.04 & This work\\ 
FRB\,20201124A & 0.0982 & 2021 Apr 20  & 29.7 & $40.6^{+0.16}_{-0.16}$ & 38.2 & 0.45 & 13
\enddata
\tablecomments{X-ray luminosity limits and detections correspond to the host center. Limits at the FRB position for each source are given in Section~{\ref{sec:obs}}. Limits correspond to 3$\sigma$.\\
$^{a}$ The total exposure time was spread across 4 epochs spanning 2015 Nov 23 - 2017 Nov 01.\\
$^{b}$ The total exposure time was spread across 2 epochs spanning 2019 Dec 03 - 2019 Dec 18.\\
$^{c}$ The total exposure time was spread across 58 epochs spanning 2000 May 07 - 2022 Jun 04.\\
$^{d}$ The total exposure time was spread across 2 epochs spanning 2022 Apr 23 - 2022 Apr 24.\\
References: 1. \citet{Scholz2016}; 2. \citet{Chatterjee2017}, 3. \citet{Scholz2017}, 4. PI: Bogdanov, 5. \citet{Scholz2020}, 6. \citet{Marcote2020}, 7. Sydnor et al. 2023, 8. \citet{Niu2022}, 9. \citet{Bhandari2020}, 10. \citet{Kirsten2022}, 11. See \textit{Facilities} statement, 12. \citet{Miller2010}, 13. \citet{Piro2021}.}
\label{tab:xrayfrbs}
\end{deluxetable*}

\subsection{Constraints on an AGN in the Host Galaxy of FRB20200430A}

We obtained {\it Chandra} observations (PI: Eftekhari) of FRB\,20200430A (a non-repeating FRB detected and localized by ASKAP; \citealt{Heintz2020}) in two separate epochs with exposure times of 22.9 and 11.9 ks on UT 2022 April 23 (Obs ID: 25252) and 24 (Obs ID: 26397), respectively.
We do not blindly detect X-ray emission at or near the position of FRB\,20200430A in either epoch using \texttt{wavdetect}. We therefore generate a co-added, exposure corrected image using \texttt{merge\_obs} and perform a targeted extraction around the host center using an aperture radius of $1\arcsec$. We perform background estimation in a similar manner to FRB\,20190608B. The source is not detected in the merged event file corresponding to a $3\sigma$ upper limit on the count rate of $3.6 \times 10^{-4} \ \rm cts \ s^{-1}$ (assuming Poisson statistics; \citealt{Gehrels1986}). For $\Gamma = 2$ (typical of AGN in nearby galaxies; \citealt{Ho2008}) and $N_{\rm H,MW} = 3.5 \times 10^{20} \ \rm cm^{-2}$, the $3\sigma$ limit on the $2-10$ keV absorbed X-ray flux is $F_X < 3.2 \times 10^{-15} \ \rm erg \ cm^{-2} \ s^{-1}$, corresponding to an X-ray luminosity $L_{\rm X} < 2.8\times 10^{41} \ \rm erg \ s^{-1}$. We derive the same flux limit at the FRB position located $\sim 1\arcsec$ from the host center.

\subsection{Re-Analysis of Archival Data}
\label{re-analysis}

We also searched the positions of all localized FRBs in the {\it Chandra} archive to compile a complete list of X-ray observations for well-localized events (see Table~\ref{tab:xrayfrbs}). We find five FRBs with robust host galaxy associations with archival \textit{Chandra} data (FRBs\,20121102A, 20180916B, 20190520B, 20200120E, and 20201124A). With the exception of FRB20200120E, all of these were targeted for FRB science. For uniformity, we reanalyze the data using the above methodology. In all cases, hydrogen column densities are derived using the method of \citet{Willingale2013} and X-ray fluxes are estimated assuming a photoelectrically absorbed power-law spectrum with $\Gamma = 2$.

For FRB\,20121102A, we combine four epochs of existing \textit{Chandra} observations, including one ACIS-S exposure from 2015 November (Obs ID:18717; PI: Scholz) and three ACIS-I exposures from 2016 November and 2017 January and November (Obs ID: 19286, 19287, 19288; PI: Bogdanov) corresponding to a total exposure time of 99.99 ks. Only two photons are detected within a 1\arcsec radius aperture centered on the host position, consistent with a non-detection. The $3\sigma$ upper limit on the count rate from Poisson statistics is $< 1.09\times 10^{-4}$ cts s$^{-1}$. For $N_{\rm H,MW} = 6.3 \times 10^{21} \ \rm cm^{-2}$, the $3\sigma$ limit on the absorbed flux is $F_X < 1.2 \times 10^{-15} \ \rm erg \ cm^{-2} \ s^{-1}$ ($2-10$ keV), corresponding to an X-ray luminosity $L_{\rm X} < 1.6\times 10^{41} \ \rm erg \ s^{-1}$. We derive the same flux limit at the FRB position.

We independently analyze the two existing epochs of \textit{Chandra} observations for FRB\,20180916B following \citet{Scholz2020}. We find four counts in a 1\arcsec radius aperture centered on the FRB host position. Accounting for the background, the total source counts are $2.02 \pm 1.01$ ($0.8 \sigma$ significance), consistent with a non-detection. The $3\sigma$ Poisson upper limit on the count rate is $3.3 \times 10^{-4}$ cts s$^{-1}$. For $N_{\rm H, MW} = 6.1 \times 10^{21} \ \rm cm^{-2}$, the $3\sigma$ limit on the absorbed flux is $F_X < 3.4 \times 10^{-15} \ \rm erg \ cm^{-2} \ s^{-1}$ ($2-10$ keV), corresponding to an X-ray luminosity $L_{\rm X} < 9.9\times 10^{41} \ \rm erg \ s^{-1}$. At the FRB position (offset from the host galaxy center by $\approx 7.8$\arcsec), we find a single count in a 1\arcsec\ radius aperture, consistent with a non-detection. The $3\sigma$ Poisson upper limit on the count rate is therefore $2.7 \times 10^{-4}$ cts s$^{-1}$, corresponding to a $3\sigma$ limit on the absorbed flux of $F_X < 2.8 \times 10^{-15} \ \rm erg \ cm^{-2} \ s^{-1}$ ($2-10$ keV).

For FRB\,20190520B, we analyze a single ACIS-S 14.9 ks exposure (Obs ID: 22370; PI: Aggarwal; Sydnor et al. 2023). We do not detect any counts in a 1\arcsec\ radius aperture centered on the host center. The $3\sigma$ upper limit on the count rate is $4.4 \times 10^{-4}$ cts s$^{-1}$. For $N_{\rm H,MW} = 2.56 \times 10^{21} \ \rm cm^{-2}$, the $3\sigma$ limit on the X-ray flux is $F_X <  3.7 \times 10^{-15}\ \rm erg \ cm^{-2} \ s^{-1}$ ($2-10$ keV), corresponding to an X-ray luminosity $L_{\rm X} < 8.6\times 10^{41} \ \rm erg \ s^{-1}$. We derive the same flux limit at the FRB position.

For FRB\,20201124A, we detect 5 counts in a 1.5\arcsec\ radius aperture centered on the host position in a 29.7 ks exposure (Obs ID:25016; PI: Piro; \citealt{Piro2021}). We find a net source count rate of $(1.6 \pm 0.7) \times 10^{-4}$ cts s$^{-1}$ (2.4$\sigma$ significance), consistent with a marginal detection, as reported in \citet{Piro2021}. We convert this count rate to an X-ray flux with $N_{\rm H,MW} = 4.5 \times 10^{21} \ \rm cm^{-2}$ and find $F_X = (1.6 \pm 0.6) \times 10^{-15} \ \rm erg \ cm^{-2} \ s^{-1}$ ($2-10$ keV). At the distance of FRB\,20201124A ($z = 0.0982$), the corresponding X-ray luminosity is $L_{\rm X} = (4 \pm 1.7) \times 10^{40} \ \rm erg \ s^{-1}$, consistent with the values reported in \citet{Piro2021} and a star-formation origin, as demonstrated in their work. Given the small spatial offset between the FRB and the host center ($0.7\arcsec$), we adopt the host galaxy X-ray flux as an upper limit for X-ray emission at the FRB position.

In the case of FRB\,20200120E (located in M81, which has been observed extensively at X-ray wavelengths), we separately compile and merge all existing \textit{Chandra} ACIS observations that cover the host and FRB position\footnote{Details of the observations included in each merged event file are given in the \textit{Facilities} statement at the end of the paper.} using the python code \texttt{superchandra.py} \citep{superchandra}, which allows us to easily query and analyze all \textit{Chandra} observations at a given set of input coordinates. We note that none of the ACIS observations comprising our merged event files were targeted for FRB science. We derive an absorbed X-ray flux at the host center (15\arcsec\ aperture radius) of $F_X = (1.52 \pm 0.01) \times 10^{-11} \ \rm erg \ cm^{-2} \ s^{-1}$ ($2 - 10$ keV) for $N_{\rm H,MW} = 4 \times 10^{20} \ \rm cm^{-2}$, corresponding to an X-ray luminosity $L_{\rm X} = (2.36 \pm 0.02) \times 10^{40} \ \rm erg \ s^{-1}$. Here, $N_{\rm H,MW}$ is derived with the \texttt{superchandra.py} script using the known linear relation between the line-of-sight optical extinction $A_V$ \citep{Schlafly2011} and the hydrogen column density \citep{Guver2009}. We do not detect any counts in a 1\arcsec radius aperture at the FRB position, corresponding to a $3\sigma$ limit on the absorbed X-ray flux of $F_X < 8.6 \times 10^{-15} \ \rm erg \ cm^{-2} \ s^{-1}$ ($2 - 10$ keV) (see also \citealt{Pearlman2023}).

\subsection{Radio Continuum Observations}

We obtained radio observations of FRBs 20200430A (20A-157; PI: Bhandari) and 20190608B (22B-115; PI: Eftekhari) with the VLA at 6 GHz (C-band) and 22 GHz (K-band) on UT 2020 September 4 and UT 2022 April 24 respectively. The field of FRB\,20200430A was observed in B-configuration with a total on-source integration time of $\sim$1\,hr. FRB\,20190608B was observed in A-configuration with a total on-source integration time of 56\,min. Both observations utilized the 3-bit samplers providing the full 4 GHz of bandwidth at C-band and 8 GHz of bandwidth at K-band, not accounting for the excision of edge channels and RFI. For FRB\,20200430A, we used 3C286 for bandpass and flux density calibration and J1504+1029 for complex gain calibration. For FRB\,20190608B, we used 3C147 for bandpass and flux density calibration and J2229-0832 for complex gain calibration. 

In both cases, we imaged the pipeline-calibrated measurement sets using 2048x2048 pixels at a scale of 0.2\arcsec~pixel$^{-1}$ and 0.02\arcsec~pixel$^{-1}$ at 6 and 22 GHz, respectively, using multi-frequency synthesis (MFS; \citealt{Sault1994}) and \textit{w}-projection with 128 planes \citep{Cornwell2008}. We do not detect radio emission at the host center or the FRB location for either FRB, corresponding to rms values of 4.32 $\mu$Jy (6 GHz) and 6.14 $\mu$Jy (22 GHz) for FRB\,20200430A and FRB\,20190608B, respectively.

We note that FRB\,20190608B was observed previously with the Australia Telescope Compact Array (ATCA) under project code C3211 (PI: Shannon) and with the VLA (19A-121; PI: Bhandari) at $\sim 6$ GHz \citep{Bhandari2020}. A radio source is marginally detected in the ATCA observations with a flux density of $\sim 65 \pm 15 \ \mu$Jy at 5.5 GHz. The VLA observations, taken in A-configuration, revealed resolved radio continuum emission with a peak brightness of $\sim 27\ \mu Jy \ \rm beam^{-1}$ near the center of the host. The inferred star formation rate (SFR) from the 5.5 GHz ATCA detection of $3.9^{+1.0}_{-0.9} \ \rm M_{\odot} \ yr^{-1}$ (using the relation of \citealt{Greiner2016}) is below the optically inferred SFR of $7.03^{+1.43}_{-1.15} \rm \ M_{\odot} \ yr^{-1}$ \citep{Gordon2023}, but we note that standard relations between radio flux and star formation result in differences of order a factor of 2 (e.g., \citealt{Yun2002,Murphy2011}). Moreover, given that the emission spans a range of spatial scales (as shown by comparing the peak brightness seen in the ATCA and VLA images), we emphasize that the radio flux densities should be regarded with caution. Indeed, for SFR$_{\rm radio}$ = $3.9 \ \rm M_{\odot} \ yr^{-1}$, the expected flux density at 22 GHz assuming a spectral index $\alpha=0.7$ is $\sim 24 \ \rm \mu$Jy, which is above our $3\sigma$ limit of $18\ \mu$Jy. This ``missing flux'' at higher frequencies is consistent with the higher angular resolution in our VLA 22 GHz observations ($\sim 0.1\arcsec$) relative to the ATCA 5.5 GHz observations ($20\arcsec \times 2\arcsec$). Thus, we conclude that the most likely origin for the extended radio emission is star formation within the host galaxy.

We further supplement our sample of seven FRB hosts with radio continuum observations from the literature. This includes radio detections of the persistent radio sources (PRS) associated with FRB\,20121102A ($F_{\rm 6 \ GHz} = 203 \rm \ \mu Jy$; \citealt{Chatterjee2017}) and FRB\,20190520B ($F_{\rm 3 \ GHz} = 202 \rm \ \mu Jy$; \citealt{Niu2022}); radio detections of the host galaxy of FRB\,20201124A ($F_{\rm 6 \ GHz} = 221 \rm \ \mu Jy$; \citealt{Ravi2022}); and a radio limit for FRB\,20180916B ($F_{\rm 1.6 \ GHz} < 18 \ \mu Jy$ ($3\sigma$); \citealt{Marcote2020}). For FRB\,20200120E (M81), we adopt the average flux density ($F_{\rm 8.4 GHz} = 103.2 \ \rm m Jy$) over seven epochs of observations as reported in \citet{Miller2010}. We list the radio luminosity for each source ($k$-corrected to rest-frame 5 GHz assuming $\alpha = 0.7$) in Table~\ref{tab:xrayfrbs}.

\section{Multiwavelength Properties of AGN in FRB Host Galaxies}\label{sec:AGN}

\subsection{The AGN in the Host of FRB20190608B}\label{frb190608_agn}

The host galaxy of FRB\,20190608B (SDSS J221604.90-075355.9) is a grand design SB(r)c galaxy and is among the most massive ($\approx 3.6 \times 10^{10} \ \rm M_{\odot}$) FRB hosts discovered to date \citep{Gordon2023}. While the FRB location is coincident with a knot of UV emission that is likely due to star formation \citep{Chittidi2021}, model fits to the full spectral energy distribution (SED) indicate it is one of two FRB hosts with a post-starburst star formation history in which the galaxy has undergone a recent quenching event \citep{Gordon2023}. Such galaxies commonly host supermassive black holes which are believed to regulate star formation via AGN-driven feedback \citep{MartnNavarro2018}. Here, we report the first detection of luminous X-ray emission spatially coincident with the center of the host of FRB\,20190608B. An AGN origin is consistent with the previous detection of broad H$\alpha$ emission in the optical spectrum indicative of a Type I Seyfert galaxy \citep{Jonathan2012,Chittidi2021}. As such, the host of FRB\,20190608B is the only one known to harbor a Type I AGN. Our X-ray detection provides further evidence for an AGN origin (Figure~\ref{fig:frb190608_image}).

\begin{figure}
\includegraphics[width=\columnwidth]{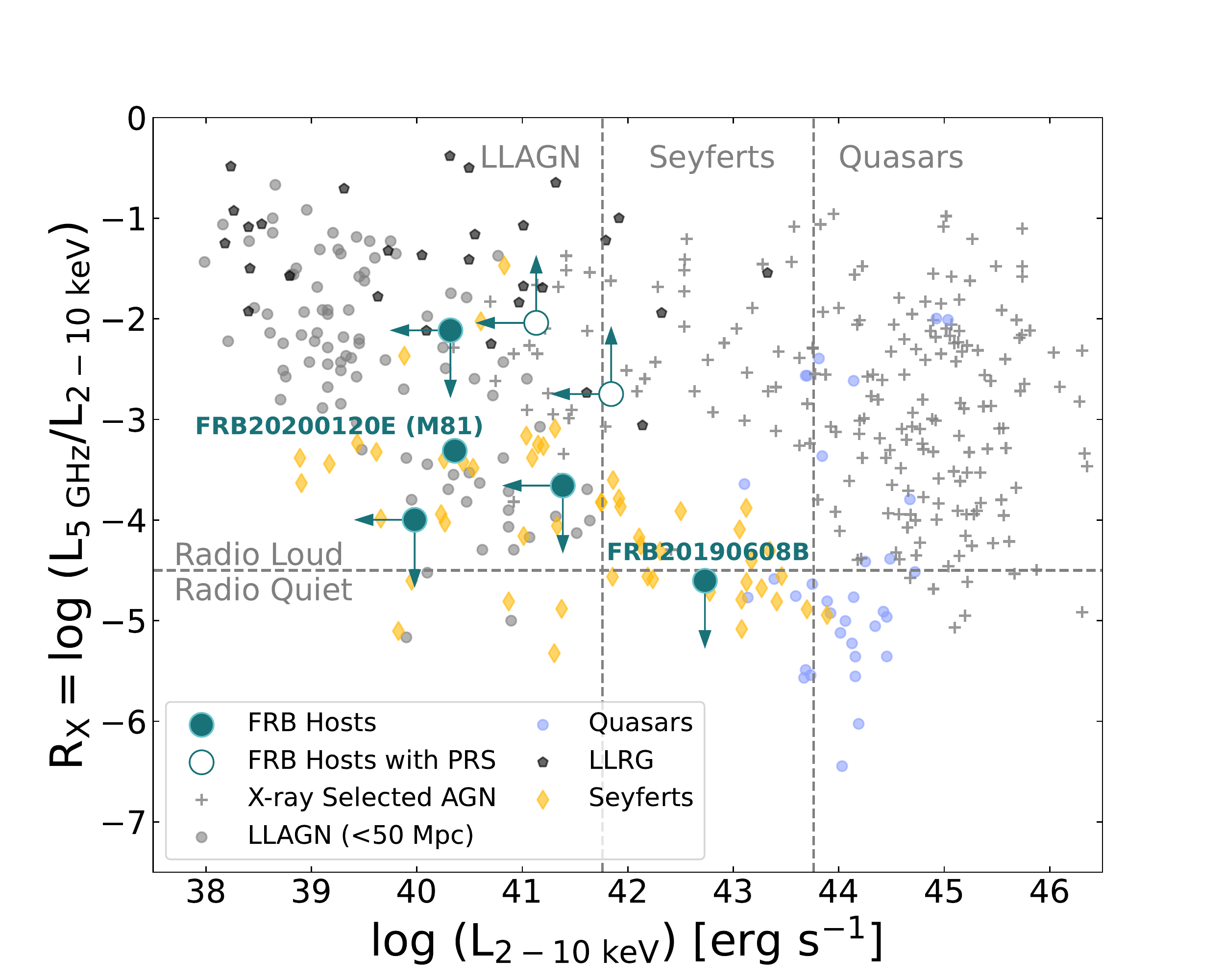}
\caption{Rest-frame radio-to-X-ray luminosity ratio ($R_X$) as a function of rest-frame X-ray luminosity ($L_X$) for FRB hosts (teal circles). Open circles correspond to FRBs with coincident PRS, where we conservatively adopt the PRS luminosity as an upper limit on the radio luminosity of an associated AGN. Shown for comparison are values for field AGN, including LLAGN \citep{She2017,Gross2019}, X-ray selected AGN \citep{Ballo2012,Lamassa2016}, Quasars \citep{Kellermann1989,Piconcelli2005,Gross2019}, Seyferts \citep{Ho2001,Terashima2003,Panessa2007}, and LLRGs; \citep{Chiaberge2005,Balmaverde2006}. Vertical dashed lines denote the nominal divisions between LLAGN and Seyferts at log($L_X \sim 41.76 \ \rm erg \ s^{-1}$) and Seyferts and Quasars at log($L_X \sim 43.76 \ \rm erg \ s^{-1}$). The horizontal dashed line at $R_X = 4.5$ roughly divides radio-loud and radio-quiet AGN.}
\label{fig:lrlx}
\end{figure}

In Figure~\ref{fig:lrlx}, we plot the rest-frame radio-to-X-ray luminosity $R_X = \rm log(L_{\rm 5 \ GHz}/L_{2 - 10 \ keV}$) as a function of rest-frame X-ray luminosity $L_X = L_{\rm 2 - 10 \ keV}$ for FRB\,20190608B, where we conservatively adopt the 5.5 GHz flux density of the host previously ascribed wholly to star formation as an upper limit on any radio emission due to an AGN (See Section~\ref{sec:obs}). For comparison, we also plot the values for several AGN samples from the literature, including low-luminosity AGN (LLAGN; $d < 50$~Mpc; \citealt{She2017,Gross2019}), X-ray selected AGN \citep{Ballo2012,Lamassa2016}, Quasars \citep{Kellermann1989,Piconcelli2005,Gross2019}, Seyferts \citep{Ho2001,Terashima2003,Panessa2007}, and low-luminosity radio galaxies (LLRGs\footnote{LLRGs are canonically unresolved, nonthermal radio cores in galaxies with optical classifications spanning Type 1 Seyferts, Type 2 Seyferts, and LINER galaxies. }; \citealt{Chiaberge2005,Balmaverde2006}). Vertical dashed lines at $\rm log(L_X) = 41.76$ and $\rm log(L_X) = 43.76 \ erg \ s^{-1}$ roughly divide AGN into three luminosity regimes corresponding to LLAGN, Seyferts, and Quasars \citep{Brusa2007}. We note however that Seyfert galaxies, particularly Type 2 Seyferts, can extend to much lower X-ray luminosities \citep{Panessa2007}, as shown in Figure~\ref{fig:lrlx}. 

Our X-ray detection of the host of FRB\,20190608B, coupled with a limit on the radio emission, places this source in the region corresponding to Seyfert galaxies, consistent with its optical classification. The value of $\rm R_X < -4.6$ is just below the nominal division between radio-loud and radio-quiet AGN ($\rm R_X = -4.5$; \citealt{Terashima2003}), although a diverse range of morphologies and properties for radio AGN challenge a simple classification \citep{Panessa2019}. The increase in $R_X$ for LLAGN with $L_X \lesssim 10^{42} \ \rm erg \ s^{-1}$ may indicate the presence of an advection-dominated accretion flow in these objects \citep{Terashima2003}, as was inferred for the LLAGN in M81 \citep{Xu2009} and consistent with its placement in this diagram.

We use the velocity dispersion at the center of the galaxy as derived from a Gaussian fit to the H$\beta$ line
($\sigma = 108$ km s$^{-1}$; \citealt{Chittidi2021}) and the $M-\sigma$ relation of \citet{Zubovas2012} for spiral galaxies to estimate the mass of the central supermassive black hole. We infer an SMBH mass of $M_{\rm BH} \sim 10^{8} \ M_{\odot}$.  For a derived stellar mass of $3.6 \times 10^{10} \ \rm M_{\odot}$, the inferred black hole mass is consistent with scaling relations between central black hole masses and galaxy stellar masses \citep{Kormendy2013,Reines2015}. Finally, adopting a bolometric correction of $L_{\rm bol} = 16 \ L_X$ \citep{Ho2008}, we find an Eddington ratio of $L_{\rm bol}/L_{\rm Edd} \approx 0.02$, typical of Seyfert galaxies \citep{Ho2008}. Moreover, this is consistent with the inferred value of $R_X < -4.6$ (and placement within the radio-quiet regime) for the host of FRB\,20190608B and reflects the fact that geometrically thin disks, which are prevalent in Seyfert galaxies, are less efficient at jet production when compared to LLAGN \citep{Terashima2003}.

\subsection{X-ray Constraints on AGN in the FRB Host Population}

Based on the \textit{Chandra} X-ray observations presented here and compiled from the literature, we find evidence for AGN in two FRB host galaxies: the host of FRB\,20190608B (Section~\ref{frb190608_agn}) and the host of FRB\,20200120E (M81), which was already known to harbor an LLAGN \citep{Miller2010}. We note that X-ray emission is also detected from the host of FRB\,20201124A but is consistent with a star formation origin \citep{Piro2021}. The detection of AGN in two out of seven FRB hosts corresponds to an occurrence rate of $\sim 29\%$. Alternatively, if we consider the hosts of repeating and non-repeating events separately, we find occupation fractions of $\sim 20\%$ and $\sim 50\%$, respectively. Conversely, the AGN fraction in the local universe is low, ranging from $0.1\%$ at $z \sim 0.2$ to $1\%$ at $z \sim 0.7$ \citep{Martini2009}. At face value, it seems that host galaxies which harbor AGN are over-represented compared to galaxies at similar redshifts. However, existing deep X-ray observations of FRB host galaxies are extremely sparse, with uncharacterized selection effects. Thus a robust assessment of the true AGN occupation fraction in FRB hosts is not yet tenable. Given that the two AGN in FRB hosts are found in both repeating and non-repeating FRB host galaxies, we observe no trend in the prevalence of AGN with FRB type, albeit based on a small sample size.

\begin{figure}
\includegraphics[width=\columnwidth]{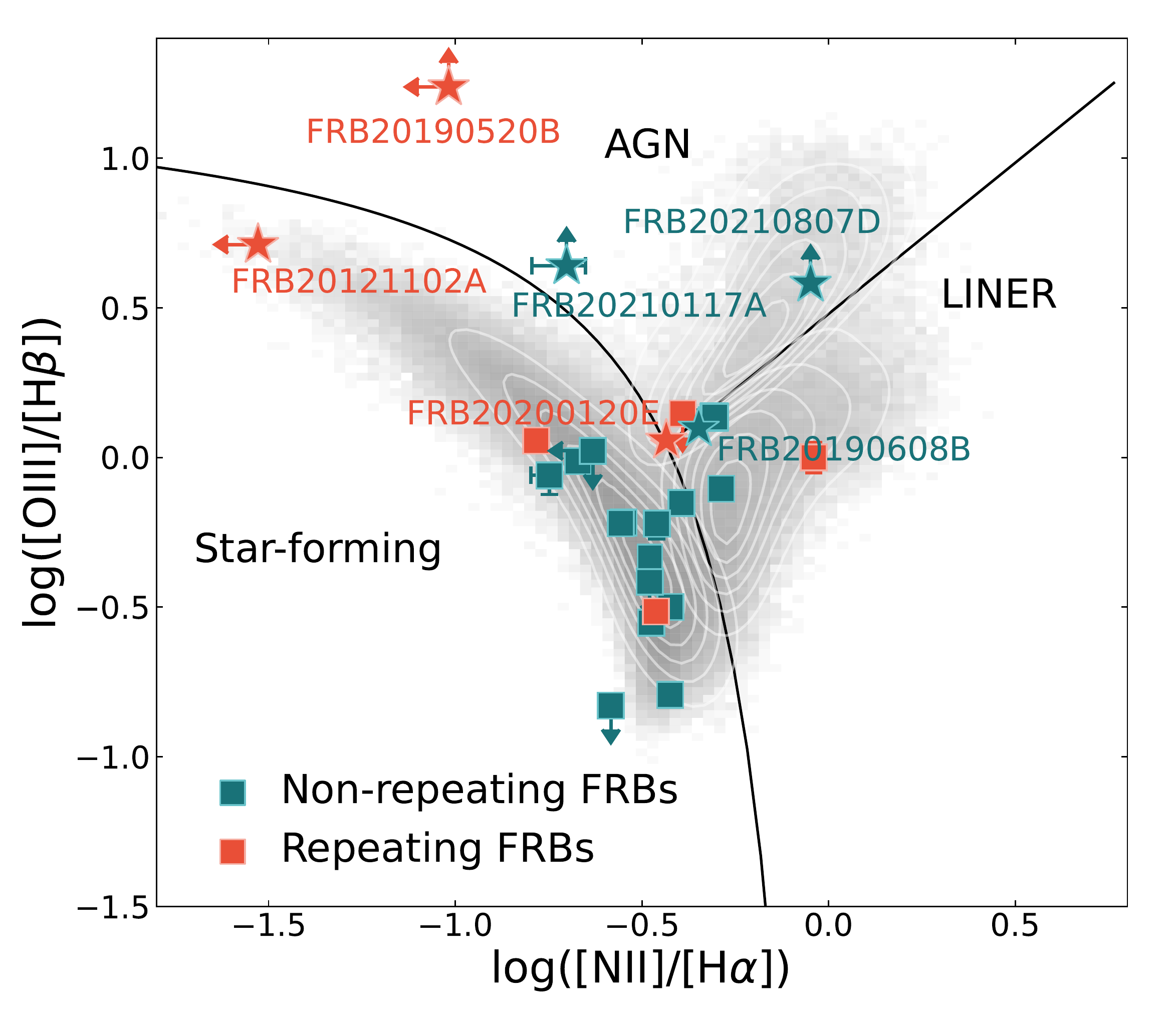}
\caption{BPT classification diagram for FRB host galaxies, including repeating (orange) and non-repeating (teal) events. Sources that are highlighted in the text are individually labeled and plotted as stars. Solid lines denote the typical demarcation between star-forming, AGN, and LINER galaxies \citep{Kauffmann2003,CidFernandes2010}. The grey-scale background corresponds to the density distribution of nearby ($0.02 < z < 0.4$) galaxies from SDSS. White contours show the KDE for each galaxy population used to model their PDFs.
}
\label{fig:bpt}
\end{figure}

To place constraints on
the presence of AGN in the absence of X-ray detections, we combine the \textit{Chandra} limits with radio observations (see Table~\ref{tab:xrayfrbs}) to plot the radio-to-X-ray luminosity $R_X$ against the X-ray luminosity $L_X$ for the remainder of sources in Figure~\ref{fig:lrlx}. We find that the X-ray limits for the majority of FRBs in our sample allow us to rule out any AGN with $L_X \lesssim 10^{41} - 10^{42} \ \rm erg \ s^{-1}$, including some LLAGN. However, we cannot rule out the low luminosity end of the LLAGN population centered around $L_{X} \sim 10^{39} \ \rm erg \ s^{-1}$, well below the sensitivity of existing X-ray observations for FRBs. For FRB\,20200430A, our X-ray limit of $L_X < 2.5 \times 10^{41} \ \rm erg \ s^{-1}$ allows us to rule out luminous quasars and most X-ray selected AGN, but we note that we cannot exclude the possibility of a Seyfert host.

Existing radio limits for FRB hosts, coupled with X-ray non-detections, correspond to radio-to-X-ray luminosity ratios of $R_X > -4$, and hence do not reach the regime of radio-quiet AGN. For FRBs\,20121102A and 20190520B, the detection of coincident PRS allows us to leverage the PRS luminosities as conservative upper limits on the radio luminosity of an associated AGN. While we cannot exclude the possibility of radio-loud LLAGN in these FRB hosts, we note that such sources are intrinsically rare in dwarf galaxies (e.g., \citealt{Reines2020}). We further note that for FRB\,20200120E (M81), and indeed all FRBs in our sample (with the exception of FRB\,20190608B), the radio and X-ray luminosities and limits are consistent with LLAGN.

\begin{deluxetable}{lccc}
\tablecolumns{4}
\tablewidth{0pt}  
\caption{2D KS Tests for FRB Host Populations}
\tablehead{
\colhead{Galaxy Type} &
\colhead{$P_{\rm KS}$ (Repeaters)} &
\colhead{$P_{\rm KS}$ (Non-Repeaters)} & 
\colhead{$P_{\rm KS}$ (All)} 
}
\startdata
All Types & 0.011 & {\bf 0.168} & {\bf 0.058}\\
SF & 0.010 & 0.011 & 0.003 \\
AGN & $<0.001$ & $<0.001$ & $<0.001$\\
LINER & 0.002 & $<0.001$ & $<0.001$\\
SF + LINER & 0.009 & {\bf 0.207} & {\bf 0.121} \\
AGN + LINER & 0.002 & $<0.001$ & $<0.001$\\
\enddata
\tablecomments{$P$-values derived from 2D KS tests comparing the FRB host populations separately (Repeaters and Non-Repeaters) and collectively (All) to different galaxy populations according to the BPT diagram. Values in bold correspond to $P_{\rm KS}>0.05$.}
\label{tab:pvalues}
\end{deluxetable}

\subsection{Optical Emission Line Diagnostics for FRB Hosts}

To determine the dominant source of ionization and distinguish between star-forming, LINER, and AGN galaxies in FRB hosts, we extract the nebular line fluxes for 22 highly secure FRB host galaxies (including 10 hosts for the first time; see Table~\ref{tab:lines}) with high quality spectra using the \texttt{Prospector} SED modeling code \citep{Johnson2021} and following the prescription and sample selection of \citet{Gordon2023}. We also include literature values for M81, the host of FRB\,20200120E \citep{Ho1996}, and FRB\,20220912A \citep{Ravi2023}.\footnote{Line fluxes for FRBs\,20200120E and 20220912A were obtained using a different method than the one we use here.}  We note that while SED fits exist for the host of FRB\,20190711A, it is not included in our sample due to poor spectral data quality which precludes robust line emission diagnostics \citep{Gordon2023}. In Figure~\ref{fig:bpt}, we plot the emission line ratios [O\,{\sc iii}]$\lambda5007$/H$\beta\lambda4861$ and [N\,{\sc ii}]$\lambda6583$/H$\alpha\lambda6563$ for FRB hosts on a BPT diagram \citep{Phillips1981} against the distribution of nearby ($0.02 < z < 0.4$) emission-line galaxies from the Sloan Digital Sky Survey (SDSS). 

Using the standard demarcation lines between star-forming, AGN, and LINER galaxies for $z \lesssim 0.3$ \citep{Kauffmann2003,CidFernandes2010}, we find that the majority of FRB hosts occupy the star-forming region, consistent with their SED classifications \citep{Gordon2023}, and are largely aligned with the locus of star-forming galaxies. Moreover, our analysis of a larger sample of hosts demonstrates that the repeating FRB hosts do not lie exclusively in the star-forming region, in contrast to previous studies \citep{Heintz2020,Bhandari2020}, and that the hosts of both repeating and non-repeating FRBs populate all three regions of the diagram roughly according to the underlying galaxy population. 

To quantify this, we compare the FRB host galaxy population, and the host populations for repeating and non-repeating events, to the distributions of star-forming, AGN, and LINER galaxies using 2D Kolmogorov-Smirnov (KS) tests to determine whether they emerge from the same underlying population. We first employ a kernel density estimation (KDE) to the SDSS sample to divide the parent population into individual galaxy classes. We next perform 2D KS tests, comparing the host populations for repeating and non-repeating events to each galaxy population, where we include all limits in the analysis. The results are summarized in Table~\ref{tab:pvalues}. 

Comparing the hosts of repeating and non-repeating FRBs collectively to all galaxy types, we find $P_{\rm KS} = 0.058$, suggesting that we cannot reject the null hypothesis that FRB hosts are drawn from the underlying galaxy population. We furthermore find that all FRB hosts are consistent with an underlying population of star-forming + LINER galaxies ($P_{\rm KS} = 0.121$), but statistically inconsistent with a population of AGN+LINER galaxies, indicating that star-forming galaxies are the dominant population. 

With only three sources in the AGN locus (one of which is a limit), we reject the null hypothesis that FRB hosts are drawn from the same population as AGN. These objects include the hosts of FRB\,20210807D, FRB\,20210117A, and FRB\,20190520B, although the latter does not exhibit detectable H$\beta$ or [N\,{\sc ii}] emission. Indeed, its location as an outlier on the BPT diagram, coupled with a dwarf host and the presence of a coincident PRS \citep{Niu2022}, bear a striking similarity to FRB\,20121102A \citep{Chatterjee2017}. However, it is worth noting that the location of a galaxy on the BPT diagram is highly sensitive to the gas-phase metallicity \citep{Kumari2021,Garg2022}. In particular, low metallicity galaxies exhibit a lower nitrogen abundance and a higher collisional excitation rate for [O\,{\sc iii}], which leads to a decrease in the [N\,{\sc ii}] flux, and an increase in the [O\,{\sc iii}] flux, respectively \citep{Garg2022}. A third dwarf host, the host of FRB\,20210117A, lies in a similar region of the diagram but hosts an apparently non-repeating FRB and lacks a persistent radio counterpart \citep{Bhandari2023}.

Hosts for the X-ray-confirmed AGN, FRBs 20190608B and 20200120E (M81), lie within the LINER region. While the precise origin of LINER emission is not settled, studies in the last few decades have shown that AGN are present in roughly $~\sim 50\%$ of LINERs \citep{Ho2001b,Satyapal2004,Dudik2005}. Here the presence of luminous X-ray emission coincident with the host nuclei for both sources points to a central AGN (as opposed to stellar ionization) as the mechanism driving the elevated levels of ionization relative to star-forming galaxies. While several other FRB hosts are located in the AGN/LINER regions, complementary X-ray constraints are lacking for these sources, with the exception of FRB\,20180916B for which X-ray observations place a fairly deep limit on the presence of an AGN (see Section~\ref{sec:obs}).

Our results are similar to previous findings in that we do find evidence for a subset of hosts residing in the LINER region and offset from the main locus of star-forming galaxies \citep{Heintz2020,Bhandari2022}. However, the larger sample presented here demonstrates that the hosts of repeating FRBs are not confined to the star-forming region, in contrast to these earlier studies. Indeed, our results show that the host galaxies for both repeating and non-repeating FRBs are statistically consistent with the full underlying galaxy population, with star-forming galaxies comprising the dominant population.

\begin{figure*}
\includegraphics[width=\columnwidth]{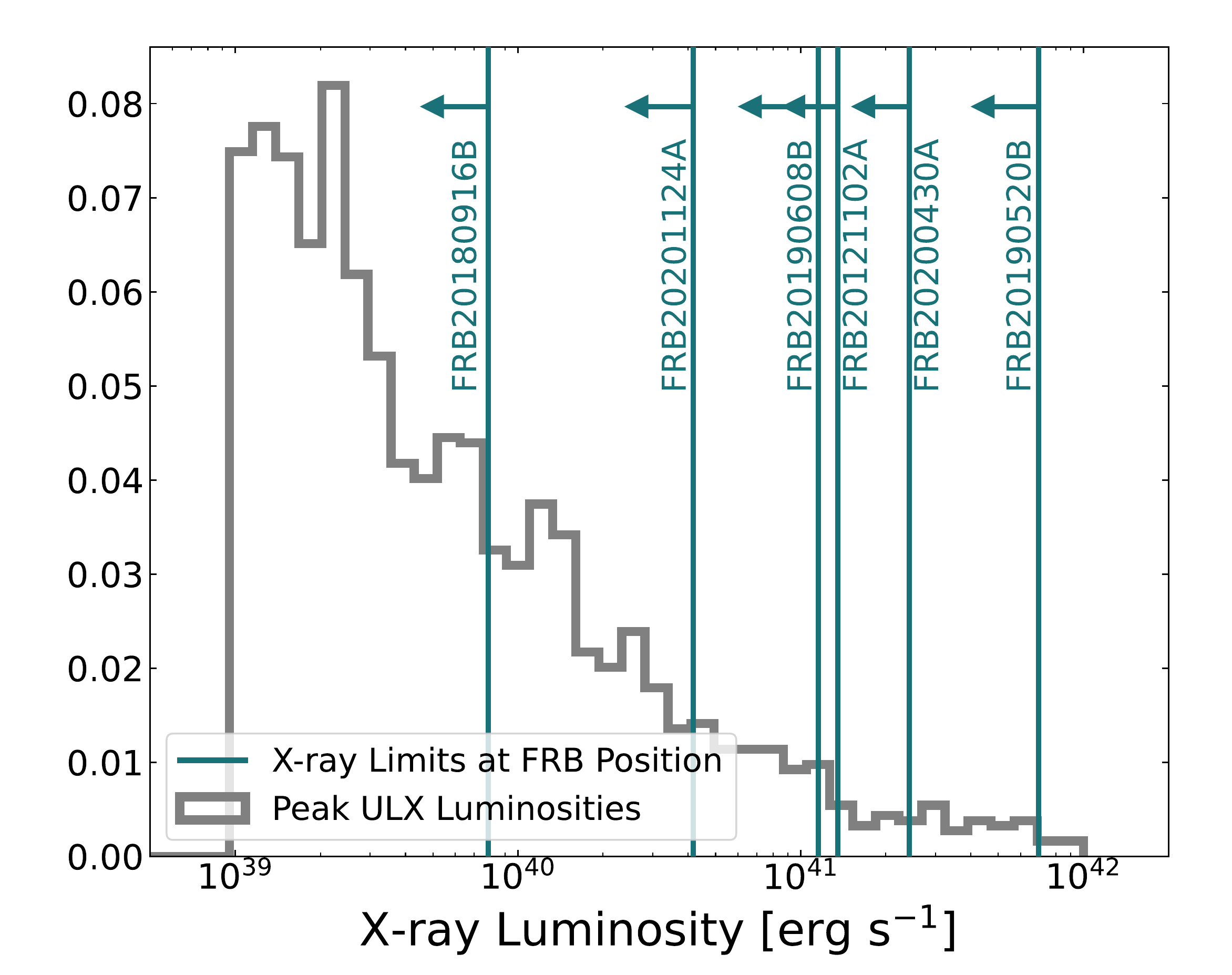}
\includegraphics[width=1.18\columnwidth]{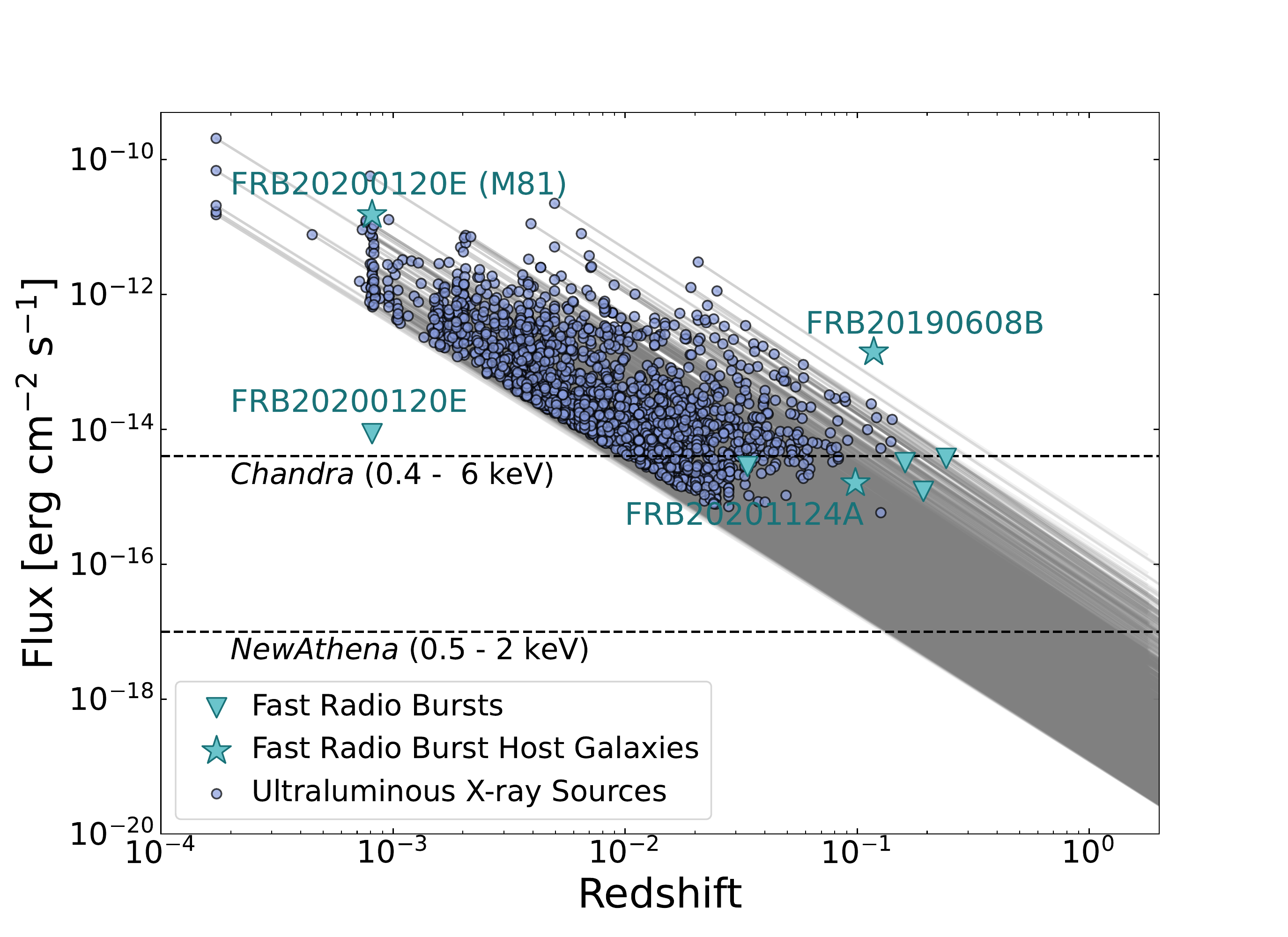}
\caption{\textbf{Left:} Upper limits on the X-ray luminosity for FRBs in our sample (vertical lines) compared to the luminosity distribution of ULXs from \citet{Walton2022}. Not shown is FRB\,20200120E, for which the X-ray luminosity limit is $L_{\rm X} \lesssim 1.3 \times 10^{37} \ \rm erg \ s^{-1}$. \textbf{Right:} X-ray limits (triangles) and detections (stars) for FRBs and FRB hosts respectively, as a function of their redshift. Shown for comparison are the fluxes of ULX (circles), where grey lines illustrate the flux scaling with redshift. The horizontal dashed lines correspond to the sensitivity of \textit{Chandra} and the planned \textit{NewAthena} X-ray observatory. While existing X-ray facilities are sensitive to ULX emission below $z\sim 0.1$, next-generation X-ray observatories such as \textit{NewAthena} will enable searches for ULX-like counterparts from FRBs out to $z\sim1$.
}
\label{fig:ulx}
\end{figure*}

\section{FRBs from Accreting Compact Objects}\label{sec:ulx}

\subsection{Searching for X-ray Counterparts}
\label{sec:xray_counterparts}

The discovery of an apparent modulation in the activity level of two luminous repeating FRB sources (FRB\,20180916B, activity modulation period of 16 days  \citep{CHIME2020_periodic}; FRB\,20121102A, activity modulation period of 160 days \citep{Rajwade2020}) has motivated progenitor theories invoking accreting stellar mass compact objects \citep{Waxman2017,Deng2021,Sridhar2021} --- most of which require super-Eddington mass-transfer (in analogy with ULXs) to explain the luminous FRB population. In this scenario, a misalignment of the black hole spin-axis with respect to the angular momentum axis of the accretion disk leads to Lens-Thirring precession of the disk and thus the jet axis \citep{Middleton2019}, driving periodicity in the FRB activity cycle on timescales of weeks to years. A prediction of this model is that FRBs --- during their active cycle --- should be accompanied by quasi-persistent X-ray counterparts. 

While the discovery of periodicity is limited to only two FRBs thus far, it is useful to use our X-ray census to constrain the ULX-like progenitor scenario. In Figure~\ref{fig:ulx} (left panel), we compare X-ray limits for the FRBs in our sample to the luminosity distribution for ULX using the multi-mission catalogue of ULXs from \citet{Walton2022}. This catalog consists of 1843 ULX sources associated with 951 host galaxies compiled from XMM-Newton, \textit{Swift}, and \textit{Chandra} X-ray source catalogs. While the sample selection is highly non-uniform and incomplete, it represents the largest ULX catalog compiled to date. 

Current ULX surveys only identify them out to average distances of 75~Mpc (maximum distance of $d_{\rm max} \approx 700$ Mpc), while our X-ray limits on a few cosmological FRBs at $z \approx 0.0337-0.1608$ are deep enough to probe the bright end of the ULX luminosity function ($\approx 10^{40}$~erg~s$^{-1}$). 
The deepest limit comes from the closest repeating FRB\,20200120E (at $3.6$ Mpc), localized to a globular cluster in the nearby M81 galaxy. The X-ray luminosity at the location of the FRB is $L_{\rm X} \lesssim 1.3 \times 10^{37} \ \rm erg \ s^{-1}$ \citep{Kirsten2022}, corresponding to $\sim 0.01 - 0.1 \ \rm L_{Edd}$ for a $10\ \rm M_{\odot}$ black hole or neutron star accretor, respectively (see also \citealt{Pearlman2023}). This is consistent with the mass transfer rate required to power bursts from FRB\,20200120E with a relatively fainter typical burst luminosity of $L_{\rm FRB}\sim10^{37}\,$erg s$^{-1}$. We note that FRB\,20200120E's localization to a globular cluster furthermore supports a connection with an old progenitor system such as accreting X-ray binaries \citep{Kirsten2022}. Conversely, the deep limit for FRB\,20180916B ($L_{\rm X} \lesssim 8 \times 10^{39} \ \rm erg \ s^{-1}$; \citealt{Scholz2020}) corresponds to $\sim 5 - 50\ \rm L_{Edd}$ \citep{Sridhar2021}. We note that the X-ray luminosity of the host of FRB\,20190608B, which is due to a Type I AGN, is well above the ULX luminosity function as expected.

Future prospects for probing the peak of the ULX luminosity function at $L_{X} \lesssim 10^{40} \ \rm erg \ s^{-1}$ will be aided by the the wealth of low-DM FRBs expected to be discovered and localized by a number of existing and planned FRB facilities (including planned upgrades). In Figure~\ref{fig:ulx} (right panel), we plot the X-ray fluxes of known ULXs \citep{Walton2022} as a function of their redshift against X-ray fluxes and limits for well-localized FRBs and their hosts. Indeed, while the present-day sample of FRB hosts is dominated by sources at $z \gtrsim 0.1$, well above the population of ULXs, a number of FRB experiments promise to enhance the detection rate of precisely localized low-DM events, pushing into the low redshift regime where ULX are primarily detected. Such experiments (and upgrades) include ASKAP with its Commensal Real-time ASKAP Fast Transients Survey (CRAFT; \citealt{Macquart2010}) COherent upgrade (CRACO) system, the More TRAnsients and Pulsars (MeerTRAP; \citealt{Sanidas2018}) project on the MeerKAT telescope, and the Deep Synoptic Array (DSA-2000; \citealt{Connor2023}) which will also increase the detection horizon for FRBs out to $z\sim 5$. The large field-of-view afforded by experiments like the Bustling Universe Radio Survey Telescope in Taiwan (BURSTT; \citealt{BURSTT2022}) and the Canadian Hydrogen Intensity Mapping FRB Experiment (CHIME/FRB; \citealt{CHIME2018}), which will achieve sub-arcsecond localizations of FRBs beginning in early 2024 thanks to the addition of outrigger telescopes to the array, will significantly increase the rate of well-localized low-DM events. Above $z \gtrsim 0.1$, where the typical ULX flux scaling drops below the \textit{Chandra} sensitivity threshold (see Figure~\ref{fig:ulx}), and as FRB experiments increase their horizon beyond the current $z\sim 1$ \citep{Ryder2022}, it will be possible with next-generation X-ray observatories, such as \textit{NewAthena} and the Advanced X-ray Imaging Satellite (AXIS), to perform productive searches for ULX-like counterparts.

\subsection{Searching for Coincident ULX-FRB Pairs}

Taking advantage of the latest ULX catalogs and the hundreds of known FRBs (albeit localized to tens of arcminutes), we search for ULXs coincident with FRBs using the multi-mission catalogue of ULXs \citep{Walton2022} described in Section~\ref{sec:xray_counterparts} and the CHIME/FRB catalog \citep{CHIME2021}.\footnote{https://www.chime-frb.ca/catalog}  The CHIME/FRB catalog includes 536 unique FRBs detected between 2018 July 25 to 2019 July 1. Among this sample, 62 events represent bursts from 18 previously reported repeating FRBs.

\begin{figure}
\includegraphics[width=\columnwidth]{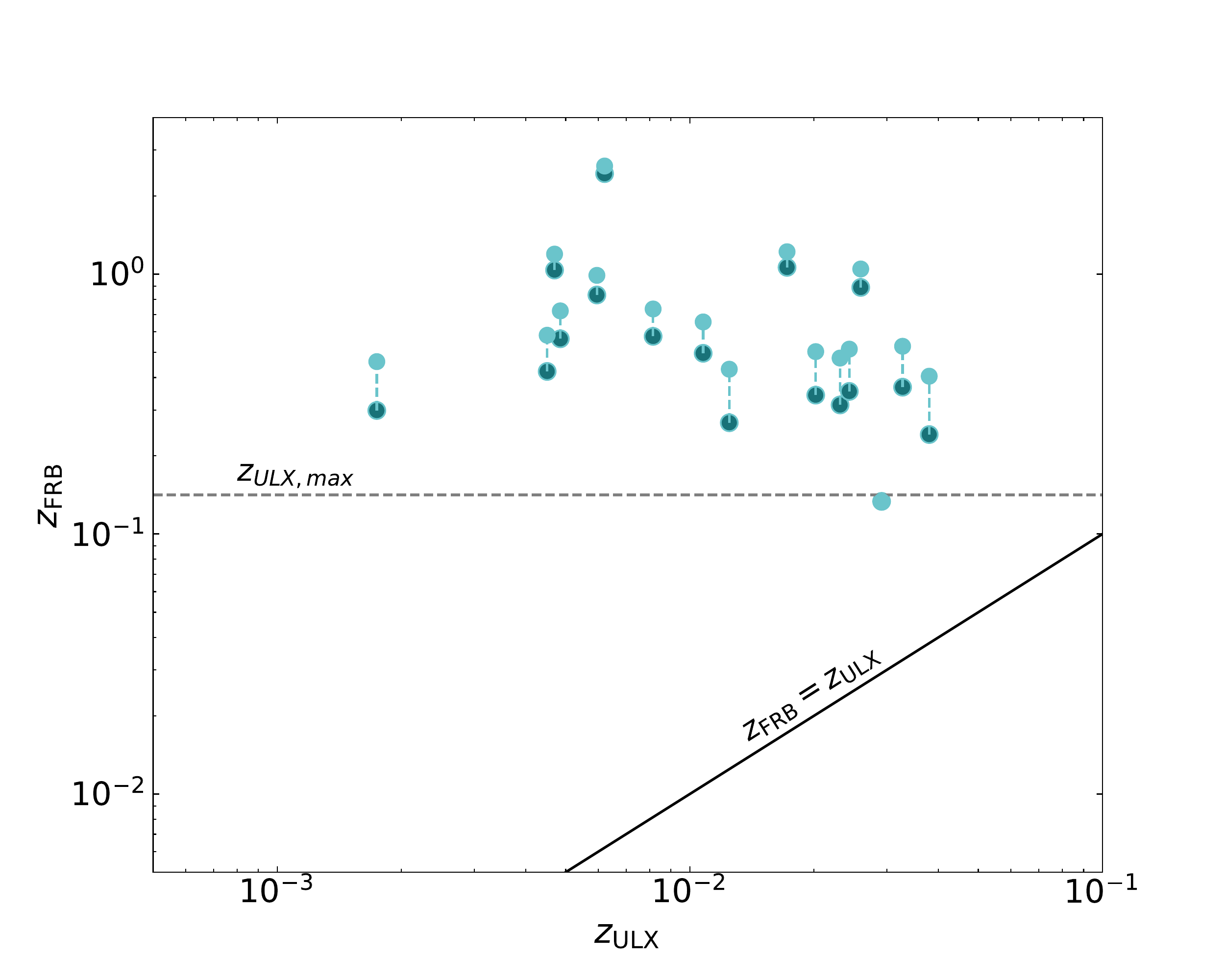}
\caption{Inferred redshifts for the CHIME/FRB Catalog 1 FRBs ($z_{\rm FRB}$) with coincident ULX in their localization regions as a function of the ULX redshift ($z_{\rm ULX}$). Dashed lines connect points with no $\rm DM_{host}$ contribution (light blue) to points with an assumed $\rm DM_{host} = 150 \ pc \ cm^{-3}$ (dark blue). The solid diagonal line indicates where $z_{\rm FRB} = z_{\rm ULX}$. The horizontal dashed line at $z \approx 0.14$ corresponds to the maximum observed redshift for a ULX from the complete ULX catalog. 
}
\label{fig:z_dist}
\end{figure}

We first search for ULXs within $3^\circ$ of a CHIME/FRB position using the FRB coordinates derived from the metadata headers \citep{CHIME2021}. Given the non-Gaussian nature of the CHIME/FRB localization regions, we do not adopt the positional uncertainties provided in the metadata headers, and instead conservatively impose a $3\sigma = 3^\circ$ localization uncertainty to account for the full extent of the CHIME localization regions. Our initial search returns 641 matches corresponding to 184 unique FRBs. We next manually check whether the matched ULXs actually lie within the FRB localization regions for each event using the HDF5 localization files. After applying this filter, we find a total of 28 ULXs spatially coincident with the localization regions for 17 FRBs. Interestingly, this includes one repeating FRB: FRB20190116A.

To assess whether the remaining FRB/ULX pairs are plausibly related, we plot the redshifts inferred from their DMs for the FRBs as a function of the ULX redshift \citep{Walton2022} in Figure~\ref{fig:z_dist}. We estimate the FRB redshifts using the Macquart relation \citep{Macquart2020} and two different values of $\rm DM_{host}$: $0 \rm \ pc \ cm^{-3}$ (shown in light blue on Figure~\ref{fig:z_dist}) and $150 \rm \ pc \ cm^{-3}$ (dark blue in Figure~\ref{fig:z_dist})\footnote{This is comparable to the median value of $\rm DM_{host} = 186^{+59}_{-48}$ estimated in \citet{James2022}.}. For the FRB DMs, we adopt the \texttt{dm\_exc\_ne2001} values from the CHIME/FRB catalog which account for the $\rm DM_{MW}$ contribution assuming the \texttt{NE2001} electron density model \citep{Cordes2002}. Given the low volumetric distance probed by existing ULX catalogs, we find that the inferred redshifts for the FRBs in our cross-matched sample are inconsistent with the ULX redshifts, assuming that the true value of $\rm DM_{\rm host}$ lies in the range $0-150 \ \rm pc \ cm^{-3}$ and ignoring scatter in the Macquart relation which may also impact our results \citep{Baptista2022,James2022,James2023,Simha2023}. In particular, inhomogeneities in the baryon distribution of the CGM may bias a DM-inferred redshift lower or higher depending on the cosmic structure along an intervening sightline. We nevertheless conclude that an association between these particular ULX-FRB pairs is unlikely.

\section{Conclusions}\label{ref:conclusions}

We have presented the first search for X-ray emission from a sample of FRB host galaxies. Our new \textit{Chandra} observations, coupled with archival \textit{Chandra} data, were used to conduct the deepest search for AGN and X-ray counterparts local to seven FRB sources. We also present an updated BPT analysis for 24 highly secure FRB hosts, presenting the nebular line properties for a large sample of FRB host galaxies for the first time. Our key results are summarized as follows:

\begin{itemize}
    \item Among the existing sample of seven well-localized FRBs with deep \textit{Chandra} observations, we find that two of the host galaxies possess nuclear X-ray sources where the emission is powered by a central AGN, including the host of FRB\,20200120E (M81), which hosts a known LLAGN. This corresponds to an AGN occupation fraction of $\sim 20\%$ and $\sim 50\%$ for repeating and non-repeating FRBs, respectively, which may point to an elevated rate of AGN in FRB hosts (when compared to field galaxies at similar redshifts). However, these results are based on a small sample size --- and uncharacterized selection effects --- warranting X-ray observations of a larger sample of hosts.

    \item Our detection of luminous ($L_X \approx 5 \times 10^{42} \ \rm erg \ s^{-1}$) X-ray emission spatially coincident with the nucleus of the host of FRB\,20190608B is consistent with its classification as a Type I Seyfert galaxy, and as such, is the only FRB host known to harbor a Type I AGN. We infer an SMBH mass of $\rm M_{\rm BH} \sim 10^{8} \ M_{\odot}$ and an Eddington ratio $L_{\rm bol}/ L_{\rm Edd} \approx 0.02$, characteristic of geometrically thin disks in Seyfert galaxies which are less efficient at jet production. We do not detect X-ray emission at the FRB position (offset by $\sim 1.6$\arcsec from the host center) and place a limit on the X-ray luminosity at the FRB location of $L_X \lesssim 10^{41} \ \rm erg \ s^{-1}$.

    \item X-ray limits for the remainder of FRBs in our sample allow us to rule out AGN with $L_X \lesssim 10^{41} - 10^{42} \ \rm erg \ s^{-1}$, including some LLAGN. However, the majority of radio limits for these sources do not reach the regime of radio-quiet AGN. 

    \item Our updated BPT analysis of 24 FRB hosts demonstrates that FRB host galaxies trace the full underlying galaxy population but are dominated by star-forming galaxies which are largely aligned with the main star-forming locus. We find that the hosts of repeating FRBs are not located exclusively in the star-forming region, however, contrary to previous studies. Similar to earlier findings, we do find evidence for an elevated fraction of hosts occurring in the LINER region.  

    \item We demonstrate that existing X-ray limits for FRBs are sufficient for probing the bright end of the ULX luminosity function ($L_X \gtrsim 10^{40} \ \rm erg \ s^{-1}$), but that the typical redshifts of FRB hosts ($z \gtrsim 0.1$) compared to the local universe distances of ULXs renders detections at the peak of the luminosity function ($L_X \lesssim 10^{40}$) challenging. Upcoming X-ray observatories like \textit{NewAthena} and AXIS will extend the detection horizon for ULX-like counterparts to FRBs out to $z \sim 1$. At the same time, upgrades to a number of FRB experiments will lend to an increased rate of low-DM FRBs which will be prime candidates for searching for ULXs at the peak of the ULX luminosity function. 

    \item Performing a cross-matching analysis between the CHIME/FRB catalog and the largest catalog of ULX sources compiled to date, we find a total of 28 ULXs spatially coincident with the localization regions for 17 FRBs. However, the DM-inferred redshifts for these FRBs imply distances $\sim 10 - 10^3$ times larger than the distances to the ULX, and hence we consider any associations unlikely. 
    
\end{itemize}

Continued X-ray observations of FRB host galaxies will critically probe the occupation fraction of AGN. Understanding the AGN fraction in FRB host galaxies, as well as pinpointing the dominant source of ionizing radiation prevalent in hosts that are optically classified as LINERS, will provide new insight into the stellar populations that should ultimately drive FRB production. Searches for X-ray counterparts will benefit most from dedicated follow-up of the most nearby events, where such observations can be most constraining.


\acknowledgments{We thank Kari A. Frank, Paul Green, Alice P. Curtin, Kristina Nyland, and Victoria Kaspi for helpful discussions.  
T.E. is supported by NASA through the NASA Hubble Fellowship grant HST-HF2-51504.001-A awarded by the Space Telescope Science Institute, which is operated by the Association of Universities for Research in Astronomy, Inc., for NASA, under contract NAS5-26555. W.F. gratefully acknowledges support by the National Science Foundation under CAREER grant No. AST-2047919, the David and Lucile Packard Foundation, the Alfred P. Sloan Foundation, and the Research Corporation for Science Advancement through Cottrell Scholar Award \#28284. N.S. acknowledges the support from NASA (grant number 80NSSC22K0332), NASA FINESST (grant number 80NSSC22K1597), and Columbia University Dean's fellowship.  S.B. is supported by a Dutch Research Council (NWO) Veni Fellowship (VI.Veni.212.058). Y.D. is supported by the National Science Foundation Graduate Research Fellowship under Grant No. DGE-1842165. A.T.D. and R.M.S. acknowledge support through Australian Research Council ARC DP DP220102305.  R.M.S. acknowledges support through Australian Research Council Future Fellowship FT190100155. 
 J.X.P., A.C.G., Y.D., W.F., T.E., N.T., and C.D.K. acknowledge support from NSF grants AST-1911140, AST-1910471
and AST-2206490 as members of the Fast and Fortunate for FRB
Follow-up team. A.B.P. is a Banting Fellow, a McGill Space Institute~(MSI) Fellow, and a Fonds de Recherche du Quebec -- Nature et Technologies~(FRQNT) postdoctoral fellow. 
}

\facilities{The VLA observations presented here were obtained as part of programs VLA/20A-157, PI: Bhandari and VLA/22B-115, PI: Eftekhari. The VLA is
operated by the National Radio Astronomy Observatory, a
facility of the National Science Foundation operated under
cooperative agreement by Associated Universities, Inc. The \textit{Chandra} ACIS data for FRB\,20200120E used in this paper can be found in the \textit{Chandra} Data Archive under the following Observation IDs for 1) the FRB position: 9540, PI: Jenkins; 4752, PI: Miller; 13728 and 14471, PI: Grise; 18051, 18052, 18817, 19685, 19992 and 19993, PI: Swartz; and 2) the host center: 735, 18047, 18048, 18049, 18050, 18051, 18052, 18053, 18054, 18817, 18875, 19685, 19688, 19981, 19982, 19991, 19992 and 19993, PI: Swartz; 5600, 5601, 6174, 6346, and 6347, PI: Canizares; 5935, 5936, 5937, 5938, 5939, 5940, 5941, 5942, 5943, 5944, 5945, 5946, 5947, 5948 and 5949, PI: Pooley; 4752, 6892, 6893, 6894, 6895, 6896, 6897, 6898, 6899, 6900 and 6901, PI: Miller; 9122, PI: Immler; 9540, 9805, PI: Jonker; 12301, PI: Dwarkadas; 13728, 14471, PI: Grise; 20624, PI: Barth; 21384, PI: Irwin; 26421, PI: Brightman.}

\software{
\texttt{astropy} \citep{astropy},
\texttt{CASA} \citep{McMullin2007},
\texttt{CIAO} (v4.13) \citep{Fruscione2006},
\texttt{matplotlib} \citep{matplotlib},
\texttt{numpy} \citep{numpy}, 
\texttt{pandas} \citep{pandas}, 
\texttt{Prospector} \citep{Johnson2021},
\texttt{SAOImageDS9} \citep{DS9},
\texttt{SExtractor} \citep{Bertin1996},
\texttt{scipy} \citep{scipy},
\texttt{superchandra} \citep{superchandra},
\texttt{xspec} \citep{Arnaud1996}. }

\begin{deluxetable*}{lcccc}
\tablecolumns{5}
\caption{Nebular Emission-line Fluxes for FRB Hosts}
\tablehead{
\colhead{FRB} & 
\colhead{H$\alpha$} &
\colhead{H$\beta$} &
\colhead{[NII]} & 
\colhead{[OIII]} \\
\colhead{} & 
\colhead{} & 
\colhead{} & 
\colhead{6584 \AA} & 
\colhead{5007 \AA} 
}  
\startdata
FRB20121102A & 2.04$^{+0.08}_{-0.09}$ & 0.78$^{+0.06}_{-0.06}$ & $<0.06$ & 4.03$^{+0.30}_{-0.30}$ \\
FRB20180301A & 3.49$^{+0.11}_{-0.10}$ & 0.99$^{+0.04}_{-0.04}$ & 0.58$^{+0.02}_{-0.02}$ &1.13 $^{+0.04}_{-0.04}$ \\
FRB20180916B & 96.75$^{+2.67}_{-2.68}$ & $<52.93$ & 39.45$^{+1.06}_{-1.04}$ & $<74.20$\\
FRB20180924B & 3.67$^{+0.08}_{-0.08}$ & 0.90$^{+0.03}_{-0.02}$ & 1.89$^{+0.04}_{-0.04}$ & 0.71$^{+0.02}_{-0.02}$\\
FRB20181112A & 1.76$^{+0.07}_{-0.07}$ & 0.50$^{+0.03}_{-0.03}$ & 0.87$^{+0.06}_{-0.08}$ & 0.69$^{+0.03}_{-0.03}$ \\%
FRB20190102C & 3.78$^{+0.29}_{-0.29}$ & 0.40$^{+0.11}_{-0.09}$ & $<0.88$ & $<0.42$\\
FRB20190520B & 0.62$^{+0.04}_{-0.04}$ & $<0.02$ & $<0.06$ & 0.40$^{+0.04}_{-0.05}$ \\%
FRB20190608B & 74.48$^{+1.59}_{-1.63}$ & 21.78$^{+0.60}_{-0.55}$ & 33.42$^{+0.68}_{-0.74}$ & 27.29$^{+0.64}_{-0.60}$\\
FRB20190611B & 1.48$^{+0.10}_{-0.10}$ & 0.31$^{+0.0.2}_{-0.02}$ & 0.26$^{+0.02}_{-0.03}$ & 0.27$^{+0.02}_{-0.04}$ \\
FRB20190714A & 9.61$^{+0.24}_{-0.22}$ & 2.01$^{+0.08}_{-0.08}$ & 3.22$^{+0.07}_{-0.08}$ & 0.57$^{+0.03}_{-0.03}$\\
FRB20191001A & 45.40$^{+0.94}_{-0.91}$ & 9.80$^{+0.43}_{-0.41}$ & 17.10$^{+0.35}_{-0.35}$ & 3.10$^{+0.09}_{-0.09}$\\
FRB20200120E$^{a}$ & 11380 & 1980 & 4180 & 2260 \\
FRB20200430A & 3.36$^{+0.13}_{-0.14}$ & 0.89$^{+0.04}_{-0.05}$ & 1.16$^{+0.04}_{-0.05}$ & 0.53$^{+0.03}_{-0.06}$\\
FRB20200906A & 14.56$^{+0.38}_{-0.35}$ & 3.14$^{+0.10}_{-0.10}$ & 3.10$^{+0.08}_{-0.08}$ & 3.06$^{+0.09}_{-0.08}$ \\
FRB20201124A & 107.93$^{+2.19}_{-2.13}$ &21.72$^{+0.60}_{-0.62}$ & 37.17$^{+0.75}_{-0.73}$ & 6.63$^{+0.18}_{-0.18}$ \\
FRB20210117A & 0.39$^{+0.02}_{-0.02}$ & \nodata & 0.08$^{+0.01}_{-0.01}$ & 0.42$^{+0.02}_{-0.03}$ \\%
FRB20210320C & 16.11$^{+0.45}_{-0.45}$ & 3.43$^{+0.16}_{-0.15}$ & 4.55$^{+0.15}_{-0.14}$ & 2.08$^{+0.08}_{-0.08}$ \\%
FRB20210410D & 2.81$^{+0.10}_{-0.11}$ & 0.32$^{+0.05}_{-0.05}$ & 0.73$^{+0.03}_{-0.04}$ & $<0.05$\\%
FRB20210807D & 22.89$^{+0.49}_{-0.55}$ & $<4.08$ & 20.44$^{+0.41}_{-0.41}$ & 15.58$^{+0.43}_{-0.39}$ \\%
FRB20211127I & 501.43$^{+11.34}_{-12.26}$ & 177.31$^{+6.18}_{-6.87}$ & 188.78$^{+4.38}_{-4.72}$ & 28.55$^{+0.98}_{-1.76}$ \\%
FRB20211203C & 11.32$^{+0.29}_{-0.27}$ & 3.73$^{+0.14}_{-0.15}$ & 3.14$^{+0.09}_{-0.08}$ & 2.25$^{+0.07}_{-0.07}$ \\%
FRB20211212A & 99.33$^{+4.38}_{-4.44}$ & $<24.71$ & 33.02$^{+1.69}_{-1.93}$ & $<9.51$ \\%
FRB20220105A & 2.36$^{+0.09}_{-0.09}$ & 0.59$^{+0.03}_{-0.03}$ & 0.95$^{+0.03}_{-0.04}$ & 0.42$^{+0.02}_{-0.02}$ \\%
\enddata
\tablecomments{Fluxes are in units of $10^{-16} \ \rm erg \ s^{-1} \ cm^{-2}$ and corrected for Galactic extinction. Not listed are the values for FRB\,20220912A, for which we use the line ratios presented in \citet{Ravi2022}. Limits correspond to $3\sigma$. \\
$^{a}$ From \citealt{Ho1996}.}
\label{tab:lines}
\end{deluxetable*}

\bibliographystyle{aasjournal}
\bibliography{references}

\end{CJK*}
\end{document}